\documentclass[preprint,12]{aastex}
\def\Dwa{$\,$\uppercase\expandafter{\romannumeral5}$\,$}

\def\sles{\lower2pt\hbox{$\buildrel {\scriptstyle <}
   \over {\scriptstyle\sim}$}}

\def\sgreat{\lower2pt\hbox{$\buildrel {\scriptstyle >}
   \over {\scriptstyle\sim}$}}
\def\sharpnull#1{}

\begin{document}

\slugcomment{\bf}
\slugcomment{for details, see \url http://www.astrophysics.arizona.edu/}

\title{Modeling Cloud Formation: Source Code}
\shorttitle{Cloud Model}

\author{Curtis S.\ Cooper\altaffilmark{1}, Jonathan I. Lunine\altaffilmark{1},
John A. Milsom\altaffilmark{2}}

\altaffiltext{1}{Department of Planetary Sciences and Lunar and Planetary
Laboratory, The University of Arizona, Tucson, AZ 85721;
                 curtis@lpl.arizona.edu, jlunine@lpl.arizona.edu}
\altaffiltext{2}{Department of Physics, The University of Arizona, Tucson,
AZ 85721;  milsom@physics.arizona.edu}

\begin{abstract}

The cloud model of Cooper et al. (2003) estimates to first-order accuracy the
cloud particle sizes typically found in brown dwarfs and planetary
atmospheres.  This model, which is one-dimensional, is based on microphysical
considerations and incorporates the results of the theories of homogeneous and
heterogeneous particle nucleation.

We have posted the source code for this cloud model for public use as a tool
for the intercomparison of planetary radiation transport models attempting to
incorporate the physics of cloud condensation.  Follow the 'Computational
Models' link from the URL above (Theoretical Astrophysics Program - University
of Arizona) for download instructions, source code, and additional
documentation.

\end{abstract}


\end{document}